# Leveraging Models to Constrain the Climates of Rocky Exoplanets

Thaddeus D. Komacek[1], Wanying Kang[2], Jacob Lustig-Yaeger[3,4], and Stephanie L. Olson[5,6]

1 Department of the Geophysical Sciences, The University of Chicago, Chicago, IL, 60637, USA

2 Department of Earth, Atmospheric, and Planetary Sciences, Massachusetts Institute of Technology, Cambridge, MA 02139, USA

3 Johns Hopkins University Applied Physics Laboratory, Laurel, MD 20723, USA

4 NASA Nexus for Exoplanet System Science, Virtual Planetary Laboratory Team, Box 351580, University of Washington, Seattle, Washington 98195, USA

5 Department of Earth, Atmospheric, and Planetary Sciences, Purdue University, West Lafayette, IN 47907, USA

6 NASA Nexus for Exoplanet System Science, Seasonal Biosignatures Team, Purdue University, West Lafayette, IN 47907, USA

In recent years, numerical models that were developed for Earth have been adapted to study exoplanetary climates to understand how the broad range of possible exoplanetary properties affects their climate state. The recent discovery and upcoming characterization of nearby rocky exoplanets opens an avenue toward understanding the processes that shape planetary climates and lead to the persistent habitability of Earth. In this review, we summarize recent advances in understanding the climate of rocky exoplanets, including their atmospheric structure, chemistry, evolution, and atmospheric and oceanic circulation. We describe current and upcoming astronomical observations that will constrain the climate of rocky exoplanets and describe how modeling tools will both inform and interpret future observations.

## 1. The promise of rocky exoplanet characterization

In the past few years, potentially rocky exoplanets in nearby stellar systems have been discovered. These planets mostly orbit M dwarf stars, which are much smaller and cooler than our Sun. As a result, the infrared signature of the atmospheres of these planets orbiting M dwarfs is much greater than that of planets orbiting Sun-like stars and provides an opportunity to understand their atmospheric composition and climate in detail using astronomical observations. In the coming decades, the observational characterization of rocky exoplanets will reveal whether the climate of Earth is unique. Upcoming observations of these planets with the *James Webb Space Telescope* and large ground-based telescopes will provide the first constraints on the climate of temperate rocky planets outside the solar system (Snellen et al. 2015; Morley et al. 2017).

Future observations of temperate rocky planets in reflected starlight with large space-based telescopes will probe whether planets with similar properties to Earth orbiting Sun-like stars ("exoEarths") are Earth-like in their atmospheric composition and climate. These observations will allow for constraints on their atmospheric composition, including biosignatures (Feng et al. 2018). Additionally, measurements of reflected light over a full rotational phase will allow for the construction of surface maps, which can determine if the planet bears both oceans and continents (Cowan & Fujii 2018).

Theory and modeling tools adapted from geoscience are critical to the search for exoplanets with climates similar to Earth. The construct of the habitable zone (HZ) (Kasting et al. 1993, see Figure 1) provides a guideline for the received stellar flux in which the climate of planets with an Earth-like atmospheric composition and silicate-weathering feedback allows for surface liquid water. If a detected planet lies at an orbital separation that is closer to its star than the inner boundary (or "inner edge") of the HZ, the surface temperature either grows almost unboundedly due to the positive feedback between surface temperature and water vapor greenhouse effect (Runaway Greenhouse limit marked in Fig. 1), or becomes so hot that water vapor molecules evaporated from the surface are transported to the upper atmosphere and escape quickly to space. Conversely, if a planet lies further from its star than the outer edge of the HZ (maximum greenhouse limit marked in Fig. 1), carbon dioxide, even given an abundant $CO_2$ reservoir replenished from volcanic outgassing, can no longer provide sufficient greenhouse warming to keep the surface from freezing over because $CO_2$ condenses out of the atmosphere. A wide range of geophysical models have predicted how the parameter space of planetary habitability depends on planetary properties such as rotation rate, surface pressure, and orbital obliquity (e.g., Yang et al. 2013; Kopparapu et al. 2014). These predictions for the range of instellations that permit surface liquid water derived from geophysical models can be tested with future observations that constrain the atmospheric composition (including abundances of carbon dioxide and water vapor) of a range of rocky exoplanets (Checlair et al. 2019). This is but one of many examples of how geophysical models for the climate of rocky planets will be both testable with astronomical observations and will be informative as to which exoplanets may be optimal to characterize in detail. In this review, we summarize recent advances and highlight key knowledge gaps in modeling the climate of rocky exoplanets in Section 2 and then describe how both basic tenets of our understanding of rocky exoplanets (e.g., the habitable zone concept) and detailed models will be tested with future observations in Section 3.

## 2. Modeling the climates of rocky exoplanets

### 2.1 Atmospheric evolution

The atmospheres of rocky exoplanets may greatly differ from that of Earth due to their formation histories, tectonic states, and atmospheric redox state as well as the spectrum and long-term evolution of their host stars. Many potentially habitable planets exist in planetary systems that are near an orbital resonance, for instance those in the TRAPPIST-1 system. This architecture indicates that the planets in the system formed further from the host star, where temperatures are more favorable for the condensation of volatile species (e.g., water and carbon-bearing molecules), before migrating inward together. As a result, planets in such systems may be more volatile-rich than Earth. Depending on the initial volatile inventory, exoplanets could have a broad range of atmosphere and surface compositions due to a wide variety of processes, discussed in detail by Schaefer et al. (this issue). Notably, some exoplanets with large initial water inventories may be aquaplanets, with fully water-covered surfaces and an absence of exposed continents. The climates of such aquaplanets have been studied in detail and may differ dramatically from Earth-like planets with continents (see Section 2.2). The wavelength distribution of incident light from the host star between ultraviolet, visible, and infrared wavelengths—and its temporal variability and long-term evolution—also plays an important role in determining the atmospheric composition and therefore the climate of exoplanets. The ultraviolet radiation received from the host star plays a key role in determining the photochemical stability of greenhouse gases (Hu et al. 2012), the extent of water dissociation and atmospheric loss to space (Cohen et al., 2015), and the formation of hazes, all of which play a key role in determining both planetary climate and astronomically observable properties (Rugheimer & Kaltenegger 2018).

The atmospheric evolution of planets orbiting small, cool stars such as M dwarfs is expected to be strongly affected by the early radiation from their host stars. M dwarf stars have a prolonged early evolutionary stage during which they are bright, opposite to the faint early evolution of our own Sun. The incident radiation that many close-in planets orbiting M dwarf stars receive at young ages is large enough that such planets are potentially unable to retain their primary atmospheres. If either the initial atmosphere of the planet largely comprised of gas from the disk in which it forms is retained or the planet builds up a secondary atmosphere through volcanism, the strong ultraviolet radiation emitted from the young M dwarf star can cause water molecules in the atmospheres of temperate planets to dissociate into atomic hydrogen (H) and hydroxyl radical ($OH^-$). The resulting H, which has minimal mean molecular weight and is only weakly bound by gravity, may subsequently escape to space (Luger & Barnes 2015). In this way, temperate rocky planets orbiting M dwarf stars can lose over ten Earth oceans by mass of water to space, desiccating their surfaces and preventing the planet from maintaining habitable conditions if the planet formed with a small water complement and the outgassing flux of water is low. The loss of the surface water complement of planets orbiting M dwarf stars would leave a detectable signature due to the resulting escape of H enhancing the relative O content of the atmosphere, permitting the buildup of tens to hundreds of bars of $O_2$ (Lustig-Yaeger et al. 2019). However, significant volcanism that leads to outgassing from the mantle may enable a second generation of surface volatiles on planets orbiting M dwarf stars.

## 2.2 Atmospheric circulation and climate

Atmospheric circulation plays a key role in mediating planetary climate and habitability and could lead to consequences that are detectable with current and future missions (see Section 3). On fast-rotating planets like Earth, the atmospheric circulation is largely driven by the contrast in received sunlight between the equator and pole and acts to transport heat poleward to reduce latitudinal temperature differences. Near the equator, this transport manifests as a Hadley circulation with upwelling of moist air in the tropics and downwelling of dry air in the subtropics that forms the deserts of the Sahara and Sonora on Earth. Rotation prevents the Hadley cell from extending to the pole, and as a result in mid-latitudes the poleward heat transport is driven by smaller-scale fluid motions. The sizes of these mid-latitude eddies are constrained by rotation, which causes faster rotating planets to have reduced poleward atmospheric heat transport compared to more slowly rotating planets. However, around much smaller and cooler stars than our Sun, rocky planets with temperate climates lie much closer to their host stars, and thus tend to be tidally locked, with a permanent dayside that always faces the star and a permanent nightside that never sees incident starlight. In this case, the atmospheric circulation is driven by the day-to-night contrast in received starlight, rather than the equator-to-pole contrast.

Models developed for the atmospheric circulation of rocky exoplanets are generally based off of numerical tools used to study the climate of Earth. These so-called general circulation models (GCMs) solve the fluid equations of motion on a sphere using approximations relevant to the study of planets with relatively thin atmospheres with slow vertical motions, coupled to a scheme that computes the radiative heating and cooling rates. GCMs study the climate of the planet at a specific epoch, and as a result require inputs from one-dimensional planetary evolution models or assumptions about the atmospheric composition, surface albedo, and continent distribution. To adapt Earth-based models for the study of rocky exoplanets, they are often removed of Earth-specific components and are modified to apply throughout the range of incident stellar radiation and atmospheric composition relevant for rocky planets.

Over the past decade, a wide range of work has studied how the climate (in particular, the spatial temperature variation and the characteristics of atmospheric circulation) and habitability of rocky exoplanets depends on planetary parameters, such as the rotation rate, surface pressure, and planetary mass (see Shields 2019 for a recent review). These studies can broadly be split into two categories: studies of the climate of fast-rotating exoEarths that can be found around Sun-like stars, and studies of tidally locked rocky planets that can typically be found close-in to small, cool M dwarf stars. The simulated climate of an exoEarth is compared to that of an Earth-sized planet orbiting an M dwarf star in Figure 2. The temperature and cloud patterns of the fast-rotating exoEarth are roughly symmetric with longitude due to the fast planetary rotation. To the contrary, the temperature and cloud patterns of the tidally-locked planet orbiting an M dwarf star have large contrasts between the dayside and nightside due to the large contrast in received stellar radiation.

## 2.3 Modeling the boundaries of the habitable zone

ExoEarths have a broad range of possible climates because planetary parameters such as rotation rate, surface pressure, surface gravity, obliquity, and incident stellar flux can greatly affect their atmospheres and potential for hosting surface liquid water. Each of these planetary properties affect planetary climate and as a result change the width and location of the HZ. For example, exoEarths that have atmospheric $CO_2$ and moderately higher surface pressure than Earth are expected to have hotter surfaces due to the pressure broadening of $CO_2$ enhancing its greenhouse effect (Goldblatt et al. 2009), but at even higher surface pressures atmospheric scattering can reverse this trend. Thus, increasing the amount of $CO_2$ can only extend the outer edge of the HZ further from the host star until the scattering of incident starlight by $CO_2$ overwhelms greenhouse warming. In an atmosphere with condensible components such as water vapor, increasing gravity can squeeze the condensible gas out of the atmosphere, reducing the mixing ratio of water and helping the planet preserve surface liquid water. Meanwhile, less water vapor in the atmosphere also weakens the greenhouse effect, which can cool the surface of planets near the inner edge of the HZ and further help preserve liquid water on their surface (Thomson & Vallis 2019). Increasing gravity increases the ability of the planet to cool to space, enabling more massive (and higher gravity) planets to remain habitable at higher values of stellar radiation than less massive planets (Kopparapu et al. 2014). Even the obliquity of a planet can play a key role in its habitability, as planets with larger obliquity than Earth could become warmer and may not have an atmospheric cold trap that prevents water on Earth from reaching the upper atmosphere and escaping to space (Kang 2019). As a result, high obliquity may push the outer edge of the HZ further from the star, but increased vulnerability to water loss may narrow the inner edge of the HZ.

Similar to exoEarths, the climate of rocky planets orbiting M dwarfs is strongly dependent on their planetary properties. Tidally locked planets can lie in a broad range of dynamical regimes dependent on their orbital period (which is equal to their rotation period). Notably, clouds driven by deep convection on the permanent dayside of the planet can strongly reflect incoming stellar radiation, increasing the albedo of the planet and cooling the surface (Yang et al. 2013). This mechanism is particularly efficient for slowly rotating planets with rotation periods longer than 15 days, because they have relatively weak east-west flows and strong dayside cloud coverage, whereas fast rotating planets with rotation periods shorter than 5 days have strong east-west jets and reduced dayside cloud coverage. As a result, tidally locked planets with shorter orbital and rotation periods reach a runaway greenhouse state where surface liquid water evaporates at lower values of incident stellar radiation (further from the star) than planets with longer rotation periods (Kopparapu et al. 2017), narrowing the HZ. Similarly, planetary radius also affects the dynamical regime of a planet, with larger planets (with the same rotation period) acting as faster rotators in a dynamical sense due to the increased ratio of the planetary size to atmospheric length scales. However, increasing radius also increases planetary mass, resulting in a net increase in surface gravity and thus a decrease in the greenhouse effect (Kopparapu et al. 2014). This trade-off enables more massive rocky planets to retain surface liquid water closer to their M-star hosts than smaller planets.

## 2.4 Coupling the ocean and atmosphere

The importance of studying the ocean circulation of exoplanets, or "exo-oceanography," for our understanding of planetary climate has only recently been appreciated but is not surprising given that the oceans interact with Earth's climate system in several ways. For example, the high heat capacity of seawater mutes seasonal temperature variations. Moreover, the oceans play a key role in equator-to-pole heat transport. This poleward transport of warm water shapes the spatial extent of sea ice, which has profound consequences for planetary albedo. The dynamic motions of sea ice, including internal flow and passive drift, further modulate climate via albedo effects. In some cases, these dynamic phenomena may favor global glaciation on rapidly rotating planets like Earth (Yue & Yang 2020).

Ocean circulation is also expected to affect the climate of tidally locked rocky planets with open ocean on the dayside of the planet. Without ocean circulation, it is expected that rocky planets orbiting M dwarf stars exist in an "eyeball" state (Pierrehumbert 2011) with the hottest temperatures located on the dayside where the Sun is directly overhead. Ocean heat transport modifies the surface temperature pattern due to the propagation of large-scale waves, leading to a so-called "lobster" state with the hottest temperatures shifted from the substellar point (Hu & Yang 2014). The net result of this circulation is to warm the planet in the global average by amplifying heat transport to the night side and muting the large day-to-night temperature contrast that would otherwise be maintained. However, continents are potential obstacles that obstruct and modify ocean circulation patterns. If a large continent was positioned on the dayside of a tidally locked planet, the effects of ocean heat transport would be limited and ocean circulation would have only a minor effect on global climate (Salazar et al. 2020).

Ocean composition also affects planetary climate. The salinity of seawater is a particularly important climate consideration that is ultimately inseparable from ocean dynamics because the salt content of seawater strongly influences the density structure of the ocean, ocean circulation patterns, and ocean heat transport. However, the importance of seawater chemistry for climate is not limited to its effects on dynamics. For example, ocean salinity affects the solubility (and thus atmospheric abundance) of important greenhouse gases. At the same time, salinity influences evaporation and the abundance of water vapor, itself a potent greenhouse agent, and sea salt aerosols serve as cloud condensation nuclei (CCN). Finally, salt depresses the freezing point of seawater, inhibiting the growth of sea ice and modulating planetary albedo. All of these salinity effects conspire to indirectly influence planetary climate (Olson et al. 2020). As shown in Figure 3, the sea ice coverage of a planet with an Earth-like size and atmospheric composition decreases with increasing salinity, leading to warmer global-averaged temperatures. These relationships may play out differently on tidally locked M-star planets because the ice-albedo feedback is weakened due to the lower reflectivity of ice at the longer wavelengths that characterize M star light relative to light from our Sun (Shields 2019).

## 3. Observational constraints on the climates of rocky exoplanets

### 3.1 Near-term observations

Observations are beginning to place constraints on the atmospheres of rocky exoplanets that orbit M dwarf stars much smaller and cooler than the Sun. *Hubble Space Telescope* observations of the transmission spectrum of potentially temperate planets TRAPPIST-1d, e, and f ruled out the existence of cloud-free hydrogen atmospheres on these planets (de Wit et al. 2018). A 100-hour continuous observation of the hot rocky planet LHS 3844b with the *Spitzer Space Telescope* found large-amplitude changes in the brightness of the planet with orbital phase consistent with a thin atmosphere with a surface pressure less than 10 bar (Kreidberg et al. 2019). Observations with ground-based telescopes presently cannot place detailed constraints on the atmospheres of rocky planets in the HZ, but can constrain the atmospheric composition of hotter rocky planets (Diamond-Lowe et al. 2018).

A wide range of observations will be able to probe the atmospheres and constrain the climates of rocky exoplanets in the coming decade. Secondary eclipse measurements of the dayside brightness when the planet passes behind its host star with the upcoming *James Webb Space Telescope* will inform us whether or not planets orbiting M dwarf stars possess thick atmospheres (Koll et al. 2019). More detailed observations of temperate planets orbiting M dwarfs in transmission when they pass between Earth and their host star will place constraints on atmospheric composition (Morley et al. 2017; Lustig-Yaeger et al. 2019), though high-altitude water clouds and other aerosols may obscure molecular species, as found by previous observations of gaseous exoplanets. Observations of planetary brightness over a full orbital phase can constrain how the planetary temperature structure depends on longitude and directly infer the presence of clouds, as their enhanced presence on the dayside reduces the emitted radiation from the planet near secondary eclipse (Kitzmann et al. 2011). Additionally, ground-based high spectral resolution observations with the next generation of large telescopes will enable constraints to be placed on the atmospheric composition of temperate rocky exoplanets (Snellen et al. 2015). Such ground-based high contrast and high spectral resolution observations may be the best path toward constraining the atmospheric composition of the nearby temperate and rocky planets in the next decade.

### 3.2 Characterizing exoEarths in reflected light

Future observations with large space-based telescopes will allow for the observational study of Earth-sized rocky planets at orbital separations similar to that of Earth orbiting the Sun. Current mission concepts to study these exoEarths include *LUVOIR* and *HabEx*. These missions would utilize coronagraphs and/or starshades that block the light from the host star, allowing astronomers to separate the light reflected off an exoEarth from that of its host star. This enables direct imaging of the planet itself, rather than inferring the presence of the planet through variations in starlight as is done when measuring the transmission of starlight through the planetary atmosphere.

Direct imaging observations of exoEarths in reflected light will allow for the first detailed constraints of their atmospheric composition, including key habitability indicators such as $H_2O$ that, if present, could constrain whether exoEarth climates are temperate (e.g., Feng et al. 2018). Observations over multiple rotation periods will enable the construction of surface maps of exoEarths (for a recent review, see Cowan & Fujii 2018). Retrievals of the surface brightness (see Figure 4) will allow for the determination of how common it is for exoEarths to have both oceans and exposed continents similar to Earth, which is a direct probe of tectonics – and possibly planetary habitability if the prevailing view that tectonics are essential for life is correct. Longer-baseline observations over a full planetary orbit may allow for the recognition of an active hydrologic cycle on habitable planets due to seasonal variations in cloud and/or ice cover as well as seasonality in atmospheric composition that arise from seasonal variations in OH radical production (Olson et al. 2018).

### 3.3 Testing geophysical models with exoplanet observations

The study of exoplanets promises to test current understanding of the climates and evolution of rocky planets. Many current theories and geophysical modeling techniques applied to study the climates of exoplanets are based on the study of Earth, and their applicability to exoplanets will be determined by comparing their predictions to future observations. For example, the atmospheric composition of habitable rocky exoplanets is expected to be more carbon dioxide-rich further from the host star, at lower values of incident stellar radiation. This is because the silicate-weathering feedback is expected to draw down less $CO_2$ into the mantle in colder climates, enabling significant $CO_2$ buildup from volcanic outgassing. Observations that constrain the atmospheric carbon dioxide abundance of a range of planets at different locations in the HZ can determine whether or not the silicate-weathering feedback is operational on the majority of habitable exoplanets – if it is, exoplanetary climates are likely more resilient to temperature perturbations, increasing the likelihood of long-term habitability (Checlair et al. 2019). Another avenue toward constraining whether long-term tectonic processes may operate on rocky exoplanets is through measurements of reflected starlight. These observations can constrain the atmospheric composition of rocky planets (Feng et al. 2018) and map the albedo contrast between oceans and continents of rocky exoplanets, which can inform us if rocky planets typically have Earth-like distributions of continent and ocean or are more likely to be ocean-covered aquaplanets or ocean-free dune planets (Lustig-Yaeger et al. 2018).

Future observations of rocky exoplanets will also test models developed to understand the three-dimensional atmospheric and oceanic dynamics of exoplanets. One key expectation from three-dimensional climate models of tidally locked planets orbiting M dwarf stars is that their daysides are cloudy due to vigorous convection, potentially keeping the planet temperate close-in to the host star (Yang et al. 2013). In this scenario, strong cloud cover would cause the dayside of the planet to emit the least thermal radiation to space, due to the cold temperature of the cloud tops. This would lead to an inverted shape of the curve of planetary emission over orbital phase and could be detectable through future infrared observations of rocky exoplanets. A second key expectation from climate models of tidally locked rocky planets with surface oceans is that the contrast between the dayside and nightside temperature decreases for hotter planets, due to the latent heating effect of water enhancing day-night heat transport (Kopparapu et al. 2017). This expectation can be tested by measuring the emitted flux from the nightsides of rocky exoplanets orbiting M dwarf stars with extremely sensitive future infrared observatories, for instance the *MIRECLE*, *Origins Space Telescope*, and *LIFE* mission concepts. Additionally, the thermal emitted radiation from tidally locked exoplanets with oceans is expected to be non-symmetric in longitude due to the effects of ocean heat transport (Hu & Yang 2014). High-precision observations covering the full orbit of a planet around its host star may be able to discern the effect of ocean circulation on the observable properties of rocky exoplanets.

In the near-term, atmospheric characterization with the *James Webb Space Telescope* will shed light on the basic climate properties of rocky exoplanets. Measurements of the planetary emitted thermal flux will determine whether planets orbiting M dwarf stars can host atmospheres (Koll et al. 2019), or if atmospheres are readily lost due to the high levels of UV radiation emitted over the long early evolution of small stars. If rocky planets orbiting small stars do sometimes host atmospheres, further observations can determine if they have undergone significant photodissociation of water and loss of hydrogen to space during an early runaway greenhouse phase (Luger & Barnes 2015), or if their surfaces can retain liquid water. One tell-tale signature of atmospheric water loss is significant build-up of tens of bars of $O_2$, which would be detectable for planets in the TRAPPIST-1 system with the *James Webb Space Telescope* via $O_2$-$O_2$ collision-induced absorption features (Lustig-Yaeger et al. 2019). Alternatively, such desiccated planets could harbor massive $CO_2$ atmospheres and resemble Venus (Morley et al. 2017; Lustig-Yaeger et al. 2019). If planets orbiting M dwarf stars both host atmospheres and do not undergo extreme water loss, then further detailed observations can constrain their atmospheres in detail. Future observations of rocky planets can determine their atmospheric composition and temperature structure as well as probe for biosignatures (a detailed discussion of biosignatures can be found in Rimmer et al. in this issue) such as oxygen and ozone or the presence of atmospheric chemical disequilibrium due to biological production (Lustig-Yaeger et al. 2019). A detection of atmospheric biosignatures would provide compelling evidence for long-term stability of a habitable climate. By using this stepwise approach, observations in the coming decade may be able to test whether rocky planets orbiting stars much smaller and cooler than our Sun have stable climates that can allow for the evolution of Earth-like life, if such planets are uninhabitable due to the strong high-energy radiation that they receive during their youth, or if planetary evolution produces a diverse range of planets with climate conditions beyond what we find in our Solar System.

Models for Earth's climate provide an essential starting point for understanding and predicting exoplanetary climates and the distribution of potentially habitable and inhabited environments in the universe. Current and future observations of exoplanets provide a testbed for geophysical models, broadening the context for understanding Earth. In the coming decades, models will play a key role in interpreting astronomical observations of rocky exoplanets by connecting the observable properties of exoplanets to their climate and chemistry. In this way, theoretical models originally developed to understand Earth will inform the empirical study of the nature and prevalence of habitable planets in our galaxy.

Yue W, Yang J (2020) Effect of sea-ice drift on the onset of snowball climate on rapidly rotating aqua-planets. The Astrophysical Journal Letters 898: L19-L26

# Figure captions

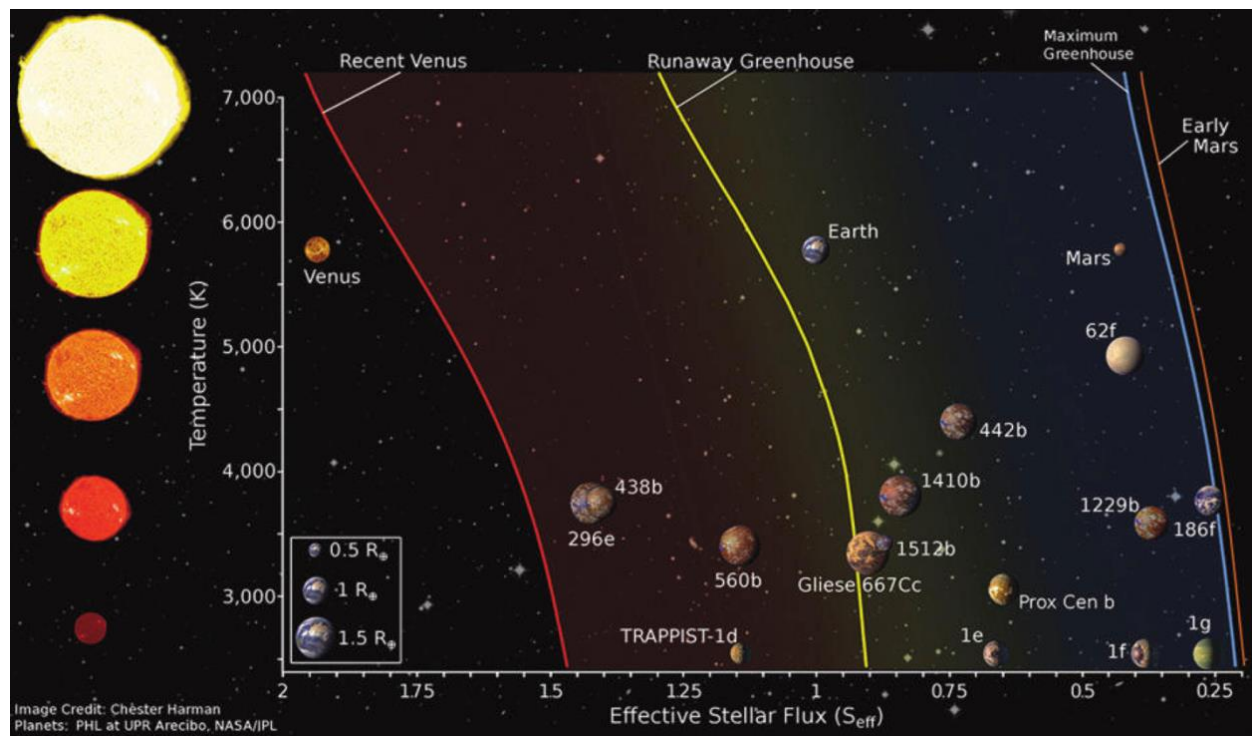

Figure 1: The positions of detected exoplanets in the habitable zone. The "recent Venus" and "early Mars" limits are derived from the potential habitability of planets in the solar system, while the "runaway greenhouse" and "maximum greenhouse" limits are derived from models of exoplanetary climate. From Schwieterman et al. (2018).

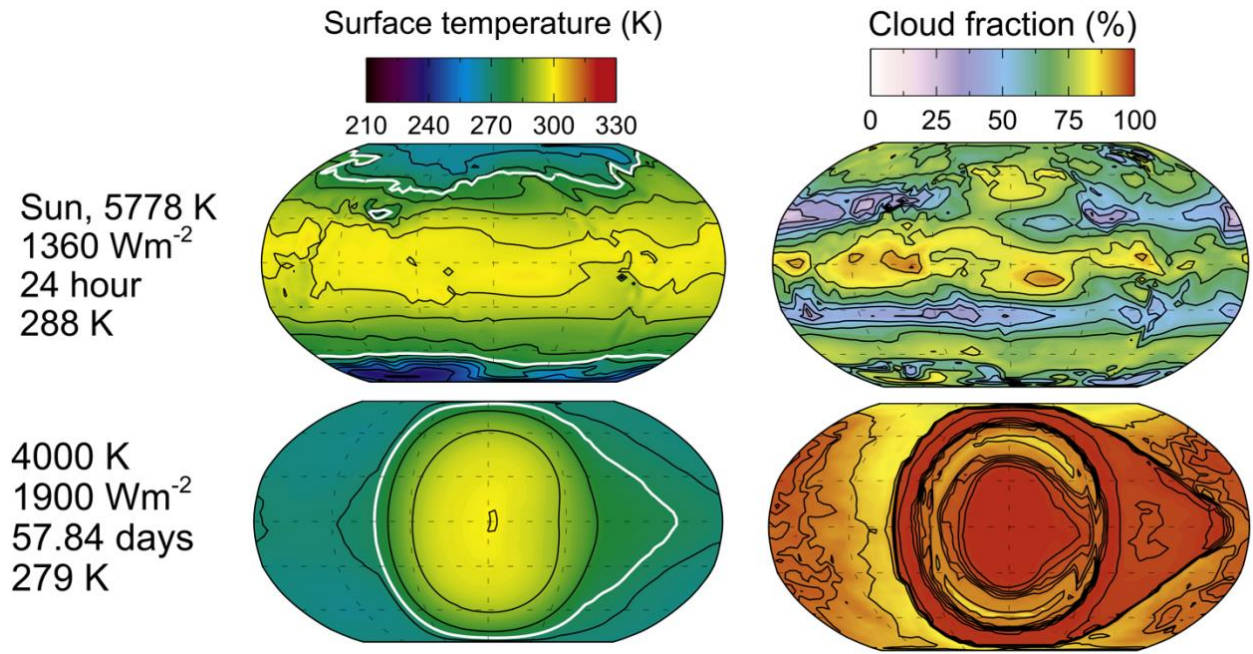

Figure 2: The surface temperature (left) and cloud coverage fraction (right) from GCM simulations of an aquaplanet orbiting a Sun-like star (top) and M dwarf star (bottom). Adapted from Kopparapu et al. (2017).

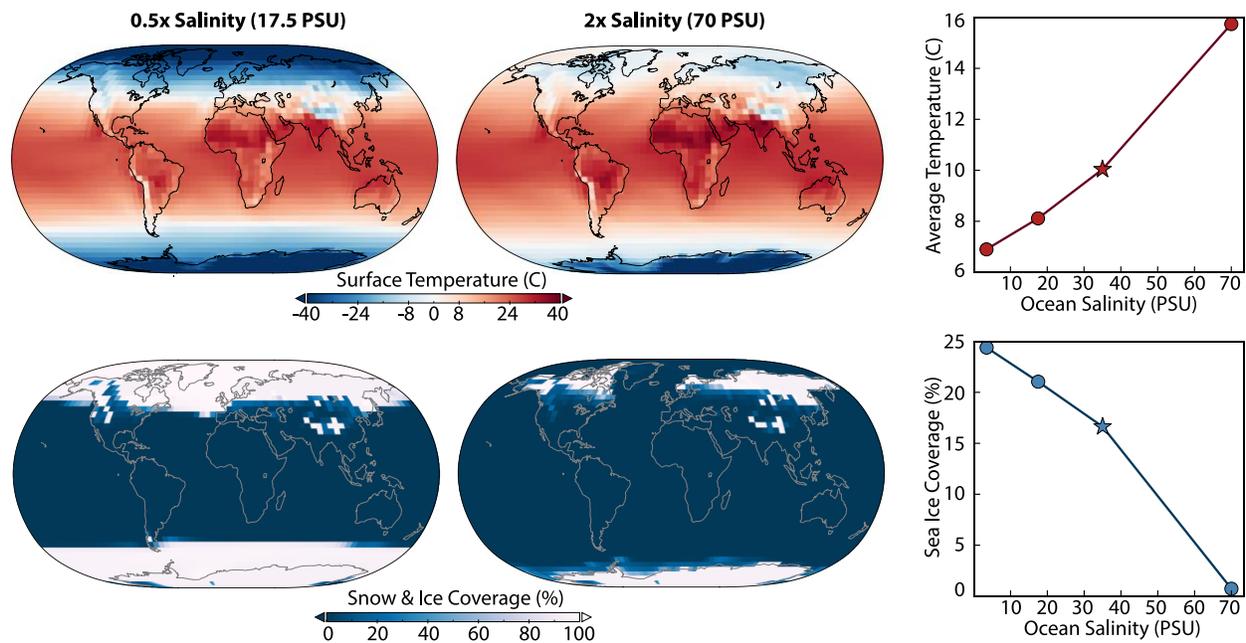

Figure 3: Maps showing surface temperature (top) and snow and ice coverage (bottom) from simulations of an Earth-like planet with lower salinity (left) and higher ocean salinity (right)

relative to present-day Earth. The plots on the right-hand side show how the globally and annually averaged temperature and sea ice coverage depend on salinity across a wider range. Modified from Olson et al. (2020).

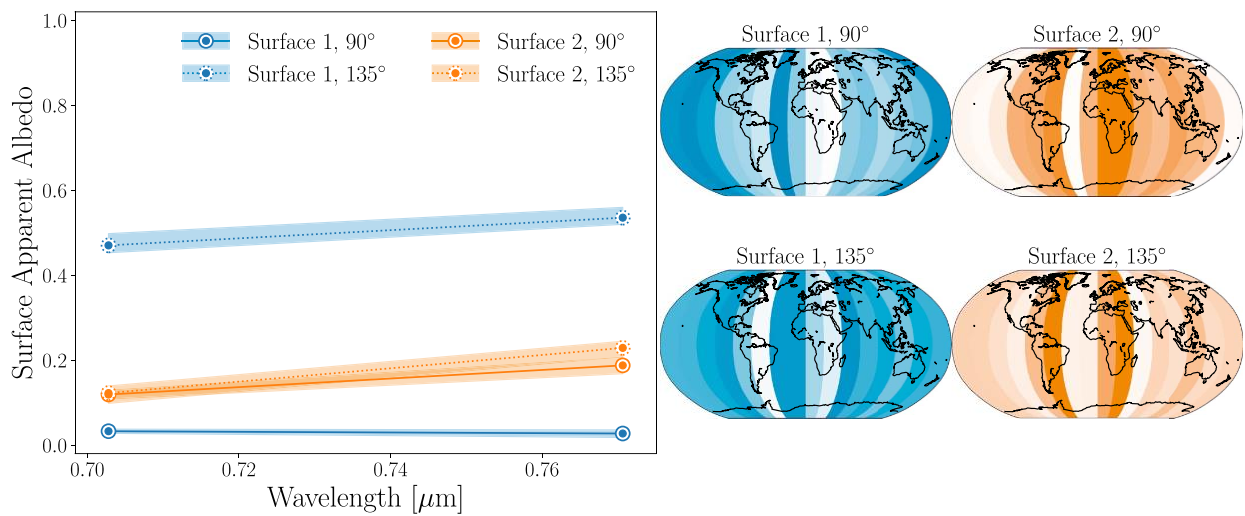

Figure 4: The apparent albedo (left) and surface maps (right) of two retrieved surfaces at orbital phases of 90 degrees and 135 degrees. The color density in the surface maps shows the retrieved surface covering area, where white corresponds to the surface not being retrieved within that slice. From Lustig-Yaeger et al. (2018).

## Acknowledgments

We thank Sarah Rugheimer and Cayman Unterborn for thoughtful comments that improved this manuscript. T.D.K. acknowledges support from the 51 Pegasi b Fellowship in Planetary Astronomy sponsored by the Heising-Simons Foundation. W.K thanks support from Lorenz-Houghton Fellowship sponsored by MIT EAPS. S.L.O acknowledges funding from the T.C. Chamberlin Postdoctoral Fellowship in the Department of Geophysical Sciences at the University of Chicago and support from the National Aeronautics and Space Administration (NASA) Exobiology and Habitable Worlds Programs. J.L.Y. acknowledges support from the Virtual Planetary Laboratory Team, a member of the NASA Nexus for Exoplanet System Science (NExSS), and the Johns Hopkins Applied Physics Laboratory's Independent Research and Development program. This work benefited from participation in the NASA NExSS research coordination network.